# Surface plasmon-polaritons in deformed graphene excited by attenuated total internal reflection

*Maksim O. Usik[†], Igor V. Bychkov[†,‡], Vladimir G. Shavrov[§], Dmitry A. Kuzmin*[†,‡],*

[†] Chelyabinsk State University, Department of Radio-Physics and Electronics, 129 Br. Kashirinykh Street, 454001 Chelyabinsk, the Russian Federation.

[‡] South Ural State University (National Research University), 76 Lenina Prospect, Chelyabinsk 454080, the Russian Federation.

[§] Kotelnikov Institute of Radio-engeneering and Electronics of RAS, 11/7 Mokhovaya Str., Moscow 125009, the Russian Federation.

ABSTRACT. In the present work we theoretically investigated the excitation of surface plasmon-polaritons (SPPs) in deformed graphene by attenuated total reflection method. We considered the Otto geometry for SPPs excitation in graphene. Efficiency of SPPs excitation strongly depends on the SPPs propagation direction. The frequency and the incident angle of the most effective excitation of SPPs strongly depend on the polarization of the incident light. Our results may open up the new possibilities for strain-induced molding flow of light at nanoscales.

I.     **Introduction**

Graphene is one of the materials which allows reaching the most possible miniaturization of nowadays devices – just a one atomic layer of thickness. Moreover, it has a lot of non-trivial physical properties ranging from mechanical to optical. For plasmonic applications, it is important that graphene carrier concentration can be tuned by chemical doping or by applying an electric field, which allows varying its electrodynamic properties from highly conductive to dielectric. This feature makes graphene a very promising material for flatland photonics and plasmonics [1-5].

For practical applications, it is highly desirable to have the effective tools for control of SPPs characteristics. Usually, this goal may be achieved by introducing some optically active materials into plasmonic structure [6-10]. In contrast to such approach, graphene may show an optical activity itself: its optical properties may be effectively manipulated by electric [11 - 13] and magnetic [14, 15] field, by topological manipulations [16, 17] or by deformations [18, 19].

Despite a large number of works devoted to electric and magnetic field manipulation by SPPs in graphene, the impact of deformations on the plasmonic properties is not investigated enough (we mean here the deformations which are not attributed to shape deformation of graphene). In this paper we investigated the excitation of surface plasmon-polaritons (SPPs) on non-elastically deformed graphene (see Figure 1). We found that such a structure shows high anisotropy of reflectance from graphene layer orientation (or orientation of deformations). Our results may pave the way for new straintronic methods of nanoscale light control.

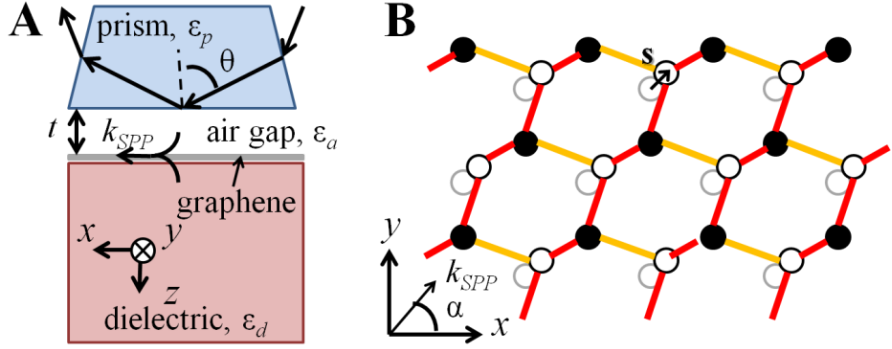

**Figure 1:** Surface plasmon-polaritons in deformed graphene may be excited in Otto configuration (**A**). Non-elastic deformation of graphene layer leads to anisotropy of propagating SPPs with respect to propagation direction (**B**).

## II. Theory

For experimental observation of propagating SPPs the number of methods may be used, for example grating structure [20] and metasurfaces [21]. Attenuated total reflection method (in Otto configuration) has been recently used for experimental observation of transverse electric polaritonic modes in graphene at infrared frequencies [22]. In the present work we will focus on this method of SPPs excitation for illustrating the anisotropic plasmonic properties of deformed graphene.

The Otto configuration of SPPs excitation method is based on the principle of attenuated total internal reflection. When the incidence angle of exciting wave is large $\theta > \theta_{cr}$ ( $\theta_{cr}$ is a critical angle for total internal reflection condition), photons from semi-infinite prisms are tunneling through the air gap and may couple to plasmons on graphene located on a semi-infinite dielectric (see Figure 1A) [23, 24].

We will suppose that the Si-based prism (dielectric constant is $\varepsilon_p = 12$) is separated from graphene on SiO$_2$ substrate (dielectric constant is $\varepsilon_d = 4$) by the air gap (dielectric constant is $\varepsilon_g = 1$) with thickness $t$. Electromagnetic wave incident from the prism under incidence angle is $\theta$. Graphene layer has a non-elastic deformations as illustrated in Fig. 1(B). Here deformation of graphene lattice is caused by the displacement **s** = ($s_x$, $s_y$) of one graphene sublattice with respect to another one. Such deformation could occur in graphene grown on substrate with an appropriate combination of lattice mismatch between the two crystals [25, 26]. Optical (but not plasmonic!) properties of graphene deformed by such a manner have been investigated [19]. It has been found that anisotropic optical absorption produces a modulation of the transmittance and of the dichroism as a function of the incident polarization angle.

The optical conductivity tensor for this modified graphene lattice has nonzero off-diagonal components and modified diagonal components [19]

$$\sigma_{xx} = \sigma_{yy} = \sigma_g(\omega)(1 \pm 2\beta\, s_y/a), \qquad (1)$$

$$\sigma_{xy} = \sigma_{yx} = 2\sigma_g(\omega)\beta\, s_x/a.$$

Here $\beta$ is the electron Gruneisen parameter, $\sigma_g(\omega)$ is the dynamical conductivity of non-deformed graphene on the frequency $\omega = 2\pi f$, $f$ is linear frequency. The conductivity of non-deformed graphene for given temperature $T$, chemical potential (or Fremi level) $\mu_{ch}$, and electrons scattering rate $\Gamma$ can be represented as interplay between the impacts of intra- and interband carriers' transitions [27]:

$$\sigma_g(\omega) = \sigma_{intra}(\omega) + \sigma_{inter}(\omega), \qquad (2)$$

$$\sigma_{intra}(\omega) = \frac{2ie^2 k_B T}{\hbar^2 \pi(\omega + i\Gamma)} \ln\left[2\cosh\left(\frac{\mu_{ch}}{2k_B T}\right)\right],$$

$$\sigma_{inter}(\omega) = \frac{e^2}{4\hbar\pi}\left[\frac{\pi}{2} + \arctan\left(\frac{\hbar\omega - 2\mu_{ch}}{2k_B T}\right) - \frac{i}{2}\ln\frac{(\hbar\omega + 2\mu_{ch})^2}{(\hbar\omega - 2\mu_{ch})^2 - (2k_B T)^2}\right].$$

In order to investigate the reflection of electromagnetic wave from the structure, one should solve Maxwell's equations with the corresponding boundary conditions at each interface. For monochromatic wave $\mathbf{E}_{\alpha\pm}, \mathbf{H}_{\alpha\pm} \sim \exp[-i\omega t + i\mathbf{k}_{\alpha\pm}\mathbf{r}]$, where ω is an angular frequency, and $\mathbf{k}_{\alpha\pm} = (k_x, k_y, \pm k_{\alpha,z})$ is a wavevector (sign "+" corresponds to the wave propagating along the z-axis, while "-" corresponds to the counter-propagating wave, α = p, a, d denotes "prism", "air", and "dielectric", consequently). With these notations, Maxwell's equation for the waves in each medium is read:

$$[\mathbf{k}_{\alpha\pm}, \mathbf{E}_{\alpha\pm}] = i\omega \mathbf{B}_{\alpha\pm}; [\mathbf{k}_{\alpha\pm}, \mathbf{H}_{\alpha\pm}] = -i\omega \mathbf{D}_{\alpha\pm};$$
$$\mathbf{B}_{\alpha\pm} = \mu_0 \mathbf{H}_{\alpha\pm}; \mathbf{D}_{\alpha\pm} = \varepsilon_0 \varepsilon_\alpha \mathbf{E}_{\alpha\pm}; \alpha = p, a, d.$$

Boundary conditions should be satisfied for total fields $(\mathbf{E}_\alpha, \mathbf{H}_\alpha) = (\mathbf{E}_{\alpha+}, \mathbf{H}_{\alpha+}) + (\mathbf{E}_{\alpha-}, \mathbf{H}_{\alpha-})$:

$$\mathbf{H}_{p,\tau}\big|_{z=0} = \mathbf{H}_{a,\tau}\big|_{z=0}; \mathbf{E}_{p,\tau}\big|_{z=0} = \mathbf{E}_{a,\tau}\big|_{z=0};$$
$$\mathbf{H}_{a,\tau}\big|_{z=t} - \mathbf{H}_{d,\tau}\big|_{z=t} = \hat{\sigma} \mathbf{E}_{a,\tau}\big|_{z=t};$$
$$\mathbf{E}_{a,\tau}\big|_{z=t} = \mathbf{E}_{d,\tau}\big|_{z=t}.$$

Graphene conductivity tensor is given by Eq. (1). These equations give the system of linear equations for amplitudes of transmitted and reflected waves in each medium. For given excitation parameters (frequency, incident angle, polarization of incident wave, deformation and chemical potential of graphene) one may calculate the reflection coefficient $R = |\mathbf{E}_{\mathbf{p}-}|^2/|\mathbf{E}_{\mathbf{p}+}|^2$ of the incident electromagnetic wave from this structure, which indicates what part of the incident wave energy passed into the SPPs excitation.

We should note that in attenuated total internal reflection configuration the electromagnetic wave in air and dielectric is decaying in z-direction, i.e. $k_{\alpha,z} = \text{Re}[k_{\alpha,z}] + i\text{Im}[k_{\alpha,z}]$, $\text{Im}[k_{\alpha,z}] \gg \text{Re}[k_{\alpha,z}]$ for α = a, d, i is imaginary unit.

### III. Results

All over the calculations we suppose the following parameters: $T = 300\ K, \mu_{ch} = 0.5\ eV, t = 2\ \mu m, \beta = 2$. Color maps of the reflectance are shown on Figure 2. One can see that in the deformed graphene the excitation pattern of surface plasmons differs from that of undeformed graphene. Excitation of SPPs by TM-incident wave is more efficient at lower frequencies. For TE-polarized incident wave excitation of SPPs occurs more significantly and at higher angles of incidence $\theta$.

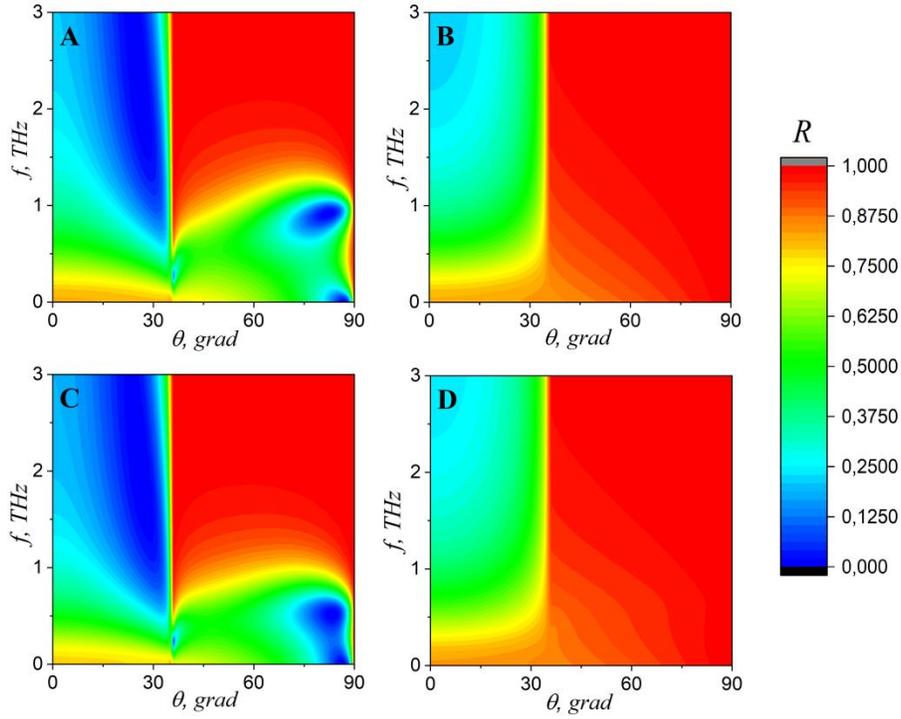

**Figure 2:** Color maps of the reflectance: a) unstrained graphene, TM-polarized incident wave; b) unstrained graphene, TE-polarized incident wave; c) deformed graphene (**s**/a = (0.04, 0.03)), TM-polarized incident wave; d) deformed graphene (**s**/a = (0.04, 0.03)), TE-polarized incident wave.

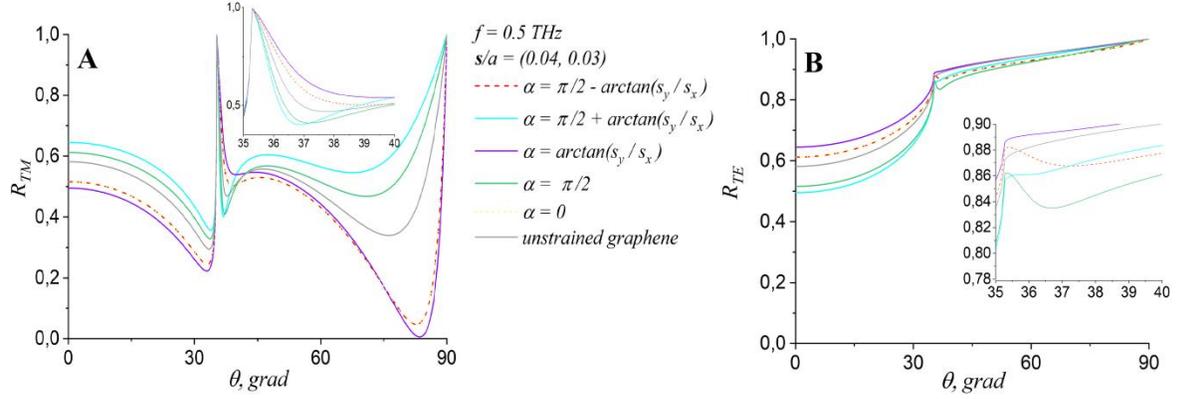

**Figure 3**: Reflectance spectrum of the system for several values of propagation angles in the x-y plane.

Fig. 3 shows that by changing the angle of propagation $\alpha$ with respect to deformation vector **s** one can control the intensity of surface plasmons excitation. There is also a coincidence of the system behavior at $\alpha = 0$ and $\alpha = \pi/2 - \arctan(s_y/s_x)$.

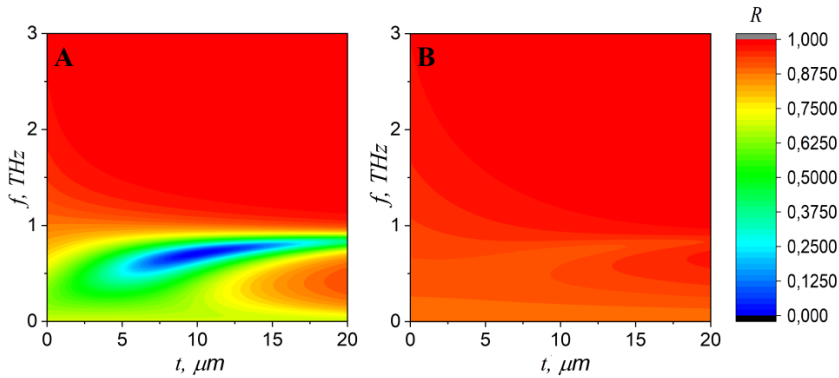

**Figure 4:** Color maps of the system's reflection coefficients: **A**) deformed graphene ($s/a = (0.04, 0.03)$), TM-polarized incident wave; **B**) deformed graphene ($s/a = (0.04, 0.03)$), TE-polarized incident wave.

Fig. 4 shows the dependence of the system behavior on the gap width. It is easy to notice that the system behavior for both TM and TE polarized incident waves are identical. The strongest excitation of surface plasmons occurs in the range up to 1 THz.

Typically for graphene-based plasmonics, the frequency range for SPPs excitation varies from THz to mid-infrared depending on graphene chemical potential [1-5]. At the frequencies corresponding to the condition of intra-band electron transitions in graphene (i.e. when $\hbar\omega \approx 2\mu_{ch}$), an imaginary part of its conductivity becomes negative (or, equivalently, its dielectric permittivity becomes positive), which means that graphene can not support any SPPs. We should note also that attenuated total reflection method is based on resonant phase matching between the incident light and propagating SPPs. This means that change of graphene chemical potential (by external gate voltage, for example) will lead to the breaking of phase matching condition and to significant change (increase) in the reflectance. SPPs excitation frequency in attenuated total internal reflection configuration depends on the part of dispersion curve which lies between the light lines of prism $\omega=ck/\varepsilon_p^{1/2}$ and dielectric $\omega=ck/\varepsilon_d^{1/2}$ [23]. Taking into account SPPs dispersion in free-standing graphene layer [4] $k \approx 2\pi\hbar^2\varepsilon_0\varepsilon_d\omega^2/(e^2\mu_{ch})$, one may calculate the maximal SPPs frequency $\omega \approx e^2\mu_{ch}/(2\pi c\hbar^2\varepsilon_0\varepsilon_d^{1/2}) \sim 1$ THz (for the parameters used all over this work). Change of material parameters of dielectric and/or graphene chemical potential allows tuning the observed part of SPPs dispersion curve.

Highly anisotropic linear SPP properties of deformed graphene considered here allow one to suppose some features for non-linear plasmonics (anisotropic second harmonic generation, for example). This problem should be solved separately in details.

**Conclusion**

The calculations have shown that deformed graphene is a good basis for the excitation of plasmons not only by TM-polarized waves, but also by TE-polarized incident waves, which is impossible to observe in an undeformed layer of graphene for THz frequency range. The frequency and angle of the most effective plasmons excitation strongly depend on the polarization of the incident electromagnetic wave. This opens up new possibilities for controlling electromagnetic radiation on nanoscale.

**Acknowledgements**

The work was financially supported in part by RFBR (16-37-00023, 16-07-00751, 16-29-14045, 17-57-150001, 19-07-00246), Act 211 of the Government of the Russian Federation (contract № 02.A03.21.0011).

AUTHOR INFORMATION

**Corresponding Authors**

*E-mail: kuzminda@csu.ru